# The Angell Plot from the Potential Energy Landscape Perspective

D. M. Zhang[a], D. Y. Sun[b,c] and X. G. Gong[a,c]

[a]Key Laboratory for Computational Physical Sciences (MOE), Institute of Computational Physics, Fudan University, Shanghai 200433, China
[b]Engineering Research Center for Nanophotonics & Advanced Instrument (MOE), School of Physics and Electronic Science, East China Normal University, 200241 Shanghai, China
[c]Shanghai Qi Zhi Institution, Shanghai 200030, P. R. China

**Abstract**

Within the scenario of the potential energy landscape (PEL), a thermodynamic model has been developed to uncover the physics behind the Angell plot. In our model, by separating the barrier distribution in PELs into a Gaussian-like and a power-law form, we obtain a general relationship between the relaxation time and the temperature. The wide range of the experimental data in the Angell plot, as well as the molecular-dynamics data, can be excellently fitted by two characteristic parameters, the effective barrier ($\omega$) and the effective width ($\sigma$) of a Gaussian-like distribution. More importantly, the fitted $\omega$ and $\sigma^2$ for all glasses are found to have a simple linear relationship within a very narrow band, and fragile and strong glasses are well separated in the $\omega$-$\sigma^2$ plot, which indicates that glassy states only appear in a specific region of the PEL.

## Introduction

The Angell plot (AP)[1], depicting the $\alpha$-relaxation time ($\tau_\alpha$) as a function of the reciprocal temperature (in units of the glass transition temperature ($T_g$)), is a very important plot in the research field of glasses. The fragility, a key concept of glasses, defined as $m = \frac{d\,log(\tau_\alpha)}{d(T_g/T)}\bigg|_{T=T_g}$, can directly be visualized in an AP. Fragility then lets us divide glasses into two classes: strong and fragile. For strong glasses, $\tau_\alpha$ can be roughly fitted by the Arrhenius equation in the range between the melting temperature and $T_g$; while for a fragile glass, $\tau_\alpha$ shows a non-Arrhenius behavior (usually referred to as super-Arrhenius).[2-4] The most essential scientific significance of the AP may be as a general classifier among various glass-forming materials.[4] For this reason, it is believed that the AP must deeply relate to the intrinsic nature of glasses.

Over past decades, great efforts have been made to look for the physical interpretation of the AP. These studies span from theoretical models to computer simulations.[5-36] Among them, the potential energy landscape (PEL) provides a *first-principles* perspective for exploring the physics behind the AP, in which the relation between the statistical properties of PELs and the fragility is thought to be a central issue. The PEL scenario of glasses comes from Goldstein's seminal paper published a half century ago.[37] Since then, it is generally believed that the PEL is indeed a very useful perspective, which can be used to investigate various glass-related problems.[38-43] Up to now, some possible connections between the PEL and the AP have been suggested. For existing models, difficulty is met when one tries to unify strong and fragile glasses within the same framework. For example, as a typical PEL-based model, the Gaussian model predicts a quadratic dependence of $\tau_\alpha$ on temperature[13] and the quadratic form well describes both strong and weak fragile glasses. However, if the quadratic form is used to fit typical fragile glasses, a negative average barrier emerges, which is obviously non-physical. Elmatad, Chandler and Garrahan (ECG) have used the quadratic form within a class of kinetically constrained consideration.[9] The ECG model does avoid the non-physical negative barrier, but it encounters other problems for strong glasses (see below for details).

In spite of great effort, the physics underlying the AP is not yet well understood, and a unified picture for both strong and fragile glasses is still missing. In this paper, we present a unified PEL model for $\alpha$-relaxation in both strong and fragile glasses using two basic facts: the existence of the barrier distribution in the PEL, and the Arrhenius relationship for a single relaxation event. The current model cannot only well establish the quantitative link between PELs and the AP, but also predicts a new general relationship to classify strong and fragile glasses.

## PEL Model

**(1) Mathematical Derivation**

It is generally recognized that the existence of metabasins (MBs) is a significant feature of PELs for glass-forming liquids.[44, 45] A basin is a region of minima in the PEL with similar potential energies. They are thought to be organized into groups, and form MBs, which have been identified in a few molecular dynamics (MD) simulations and experimental studies.[46-48] Since the $\alpha$-relaxation is related to the transition between MBs,[49, 50] we will limit our discussion to the inter-MB relaxation.

Consider a system currently staying in a MB, there are many paths connecting to other MBs as schematically shown in Fig.1. Clearly, the number of paths is closely related to the local configuration entropy. Define a function $\rho(x)$, which counts the number of paths with the barrier $x$. Hereafter $\rho(x)$ is called as the distribution of barriers (DOB). Since the system will try each path with an equal probability, a single jump can be described by the Arrhenius formula, and the average $\alpha$-relaxation time ($\tau_\alpha$) is

$$\frac{1}{\tau_\alpha} = \frac{1}{\tau_0} \int \rho(x) e^{-\beta x} dx, \quad (1)$$

where $\tau_0$ and $\beta$ are the characteristic timescale and reciprocal temperature, respectively.

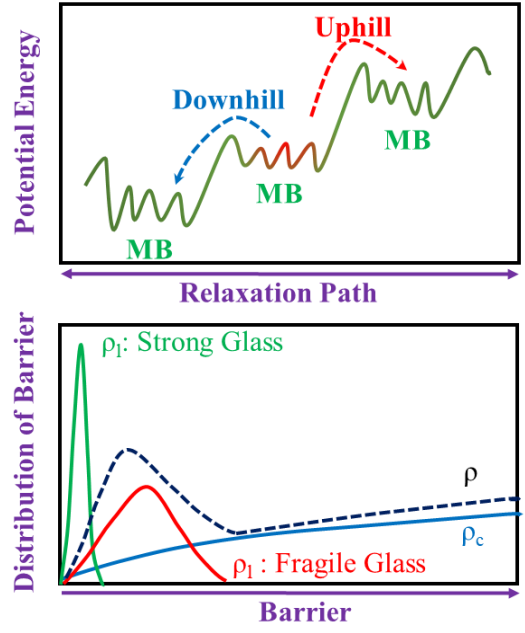

**Figure 1. Upper Panel:** Schematic plot of metabasins (MB) and possible relaxation paths. **Lower Panel:** Possible shapes of the barrier distribution. The total distribution of barriers $\rho(x)$ is divided into two parts: the distribution mainly coming from downhill relaxations $\rho_l(x)$ and from uphill relaxations $\rho_c(x)$. For a given $\rho_c(x)$ (blue solid line), the possible $\rho_l(x)$ for the strong and fragile glass are plotted in green and red, respectively.

For a given path, we label the average potential energy of the initial and final MBs as $E_i$ and $E_f$, respectively. For the **uphill relaxation** (see Fig. 1), as the difference between $E_i$ and $E_f$ is large, it is a good approximation to write $x \approx E_f - E_i$. With this approximation, the DOB for larger barriers ($\rho_c(x)$) should be $\rho_c(x) \approx \Omega(E_f) \approx \Omega(E_i + x)$, in which $\Omega(E_f)$ corresponds to the number of states with a potential energy of $E_f$. Clearly $k_B \ln(\Omega(E_f))$ is nothing but the configurational entropy at fixed potential energy $E_f$. Because the entropy is an extensive quantity, the leading term of $\Omega(E)$ may have the form, as discussed in many texts,[51] $R_c \left(\frac{x+E_i-E_0}{\mathbb{N}}\right)^{B_c \mathbb{N}}$, where $R_c$ and $B_c$ are material depended constants, and $\mathbb{N}$ is a constant proportional to the number of atoms or volume. $E_0$ is the true ground-state energy of the system (**note:** not the energy of glass at zero temperature). Finally,

$$\rho_c(x) = R_c \left(\frac{x + E_i - E_0}{\mathbb{N}}\right)^{B_c N} = R_c e^{g(x)}. \tag{2}$$

Regarding to Eq. 2, two points should be made. 1) It meets the physical requirement that entropy is an extensive quantity, and entropy will not change with the translation of energy as a whole. A similar consideration, namely the entropy being additive, has been used to analyze possible formulas for the density of states.[52-55] 2) Eq. 2 may only include the leading term, which refers to the formula in the thermodynamic limit ($N \to \infty$). For practical systems, $g(x)$ may contain other similar power terms, but this does not affect our discussion other than increasing mathematical complexity. The long tail of high energies found in some papers may relate to this kind of distribution.[19, 28, 56-58]

For **downhill relaxations**, the barrier usually does not correlate to the energy of the final states, and will distribute in a narrow region. Physically, downhill DOB should approach zero for both large and small $x$, which is schematically shown in the lower panel of Fig. 1 (blue or red lines). If $\rho_l(x)$ is used to describe the DOB of downhill relaxation, the total DOB $\rho(x) = \rho_l(x) + \rho_c(x)$.

It needs to be stressed that if the barrier $x$ is smaller or comparible to $E_f - E_i$, the above approximation for $\rho_c(x)$ may not be appropriate. Physically, $\rho_c(x)$ should be zero for $x$ less than a certain positive value, and may not be a smooth function as is Eq. 2 for small $x$. Here for mathematical convenience, we still define $\rho_c(x)$ in the range of $[0:\infty]$, and incorporate the corresponding correction into $\rho_l(x)$. With this correction, $\rho_l(x)$ and $\rho_c(x)$ do not come purely from downhill and uphill relaxation, respectively. However, we still name $\rho_l(x)$ and $\rho_c(x)$ as the DOBs for downhill and uphill relaxations, respectively.

In real materials, $\rho_l(x)$ may contain multiple maxima, in order to keep the form concise, only one dominant maximum is considered. For real systems, we can always write it as the sum of a few functions, each containing a single maximum. The final mathematical form should not change with the number of maxima. Without losing generality, we rewrite $\rho_l(x)$ as $R_l e^{-f(x)}$ with $f(x) \geq 0$. Further, we can express $f(x)$ as a polynomial, namely, $f(x) = \sum_{i=0}^{M} a_i x^i$. Based on the above assumptions,

$\rho_l(x)$ has a single maximum, as does $f(x)$. Assuming the maximum appears at $x_0$, mathematically, there will always exist $n_M$ and $n_m$ to make $A(x-x_0)^{n_M} \geq f(x) \geq a(x-x_0)^{n_m}$ true for any $x$. For the convenience of mathematical analysis and keeping major physics unchanged, we take the form $f(x) = B_l(x-\Delta)^n$, where $B_l$, $R_l$, $\Delta$, and $n$ are constants larger than zero. By properly choosing the values of $B_l$, $R_l$, $\Delta$, and $n$, $f(x)$ can roughly describe a specific system.

Considering $\rho(x) = \rho_l(x) + \rho_c(x)$, Eq. 1 becomes
$$\frac{1}{\tau_\alpha} = \frac{1}{\tau_l} + \frac{1}{\tau_c}, \tag{3}$$
with
$$\frac{1}{\tau_l} = \frac{1}{\tau_0}\int \rho_l(x) e^{-\beta x} dx = \frac{R_l}{\tau_0}\int e^{-\beta x - f(x)} dx,$$
and
$$\frac{1}{\tau_c} = \frac{1}{\tau_0}\int \rho_c(x) e^{-\beta x} dx = \frac{R_c}{\tau_0}\int e^{-\beta x + g(x)} dx.$$

Both $e^{-\beta x - f(x)}$ and $e^{-\beta x + g(x)}$ should be fast-decaying functions of $x$. Namely it is important only in the vicinity of the maximum of $-\beta x - f(x)$ and $-\beta x + g(x)$. After some mathematical deductions (see Supplemental Material for details), we reach the final results,

$$\frac{1}{\tau_l} \approx \frac{R_l}{\tau_0}\sqrt{\frac{2\pi}{f''(\bar{x})}} e^{-\frac{\Delta}{k_B T} - (n-1)B_l\left(\frac{1}{nB_l k_B T}\right)^{\frac{n}{n-1}}}, \quad \text{and} \tag{4}$$

$$\frac{1}{\tau_c} \approx \frac{R_c}{\tau_0}\sqrt{\frac{2\pi}{|g''(\bar{x})|}} e^{\frac{E_i}{k_B T} - \frac{E_0}{k_B T} + B_c N + B_c N \ln\left(\frac{B_c N}{N\beta}\right)}. \tag{5}$$

In order to further simplify the mathematics, we keep only the first two leading terms, which change fastest with temperature. In Eq. 4 and 5, besides those explicitly containing temperature, $E_i$ is also a function of temperature, which is the average potential energy in the current MB, similar to the configurational energy (total energy minus vibrational energy).[43] $E_i$ must be a monotonically increasing function of temperature, because the derivative of energies over temperature is the specific heat. In the temperature range focused on in the current work, quantum effects can be neglected, and the change of energy with temperature should be very close to linear. $E_i$ may be approximated as $\sim \epsilon_0 + \epsilon_1 T + O(T)$ (here $\epsilon_0$ and $\epsilon_1$ are constants for a given system, $\epsilon_0$ is the energy of the glasses at zero temperature). Here $O(T)$ counts those terms changing slowly with temperature. The last term in the exponential function of Eq. 5, $B_c N \ln\left(\frac{B_c N}{\mathbb{N}\beta}\right) \propto g(\bar{x})$, is proportional to the configuration entropy, since $R_c e^{g(\bar{x})}$ is the number of configurations. The dependence of entropy on temperature is much weaker than that of energy, since $T\frac{dS}{dT} = \frac{dE}{dT} = C$, the specific heat. For example, the entropy of a binary Lennard-Jones glass is about $T^{0.4}$,[59] and is similar in other glasses.[60] Considering the above arguments, Eq. 4 and 5 become

$$\frac{1}{\tau_l} \approx \frac{1}{\bar{\tau}_l} e^{-\frac{\omega_l}{k_B T} - \left(\frac{\sigma_l}{k_B T}\right)^{n_l}}, \quad \text{and} \tag{6}$$

$$\frac{1}{\tau_c} \approx \frac{1}{\bar{\tau}_c} e^{-\frac{\omega_c}{k_B T}}. \tag{7}$$

As an approximation, here the constants and the part slowly changing with temperature in Eq. 4 and 5 are collected into two new constants, $\bar{\tau}_l$ and $\bar{\tau}_c$, respectively. According to our definition, $\omega_l = \Delta > 0$, and $\omega_c = E_0 - \epsilon_0$ is the difference between the ground-state energy of systems $E_0$ and the zero-temperature energy of the glass $\epsilon_0$, obviously $\omega_c \leq 0$. Then $\sigma_l = (n-1)^{\frac{n-1}{n}} \frac{\sqrt[n]{B_l}}{n}$ measures the effective width of the downhill distribution $\rho_l$. According to our analysis in the Supplemental Material, $n_l$ must be less than or equal to 2 due to $n_l = \frac{n}{n-1} \leq 2$. $n_l = 2$ corresponds to $\rho_l(x)$ being a Gaussian distribution.

To calculate the total relaxation time $\tau_\alpha$ determined by Eq. 3, we consider three cases. First, if $\tau_l \ll \tau_c$, then $\tau_\alpha \approx \tau_l$. We have

$$\ln \tau_\alpha \approx \ln \bar{\tau}_l + \frac{\omega_l}{k_B T} + \left(\frac{\sigma_l}{k_B T}\right)^{n_l}. \tag{8.1}$$

Second, if $\tau_l \gg \tau_c$, then $\tau_\alpha \approx \tau_c$, we have

$$\ln \tau_\alpha \approx \ln \bar{\tau}_c + \frac{\omega_c}{k_B T}. \tag{8.2}$$

Third, if $\tau_l \sim \tau_c$, $\frac{1}{\tau_\alpha} = \frac{1}{\tau_l} + \frac{1}{\tau_c} \approx 2\frac{1}{\sqrt{\tau_l}}\frac{1}{\sqrt{\tau_c}}$, we have

$$\ln \tau_\alpha \approx \ln \bar{\tau}_l + \ln \bar{\tau}_c + \frac{\omega_l}{k_B T} + \left(\frac{\sigma_l}{k_B T}\right)^{n_l} + \frac{\omega_c}{k_B T}. \tag{8.3}$$

Eq. 8.3, which covers all cases by properly choosing parameters, can be further simplified as

$$\ln \tau_\alpha \approx \ln \bar{\tau}_0 + \frac{\omega}{k_B T} + \left(\frac{\sigma_l}{k_B T}\right)^n, \qquad (9)$$

where the parameters $\bar{\tau}_0$, $\omega$, $\sigma_l$ and $n \leq 2$ are material dependent, with $\omega$ and $\sigma_l$ representing the effective barrier and the effective width of the DOB, respectively. Since $n \leq 2$, the dependence of the logarithm of the relaxation time on the reciprocal temperature cannot be higher than the second power or lower than the first power.

For $\rho_l(x)$, the Gaussian distribution may be a reasonable assumption, which is also used in previous models.[13, 15-17] More importantly, some calculations also show the existence of a Gaussian-like distribution.[28, 61-65] If $\rho_l(x) \sim \exp\left[-\left(\frac{x-\Delta}{2\sigma}\right)^2\right]$ is a Gaussian distribution with the distribution width $\sigma$ and the average barrier $\Delta \geq 0$, Eq. 9 becomes

$$\ln \tau_\alpha \approx \ln \bar{\tau}_0 + \frac{\omega}{k_B T} + \left(\frac{\sigma}{k_B T}\right)^2. \qquad (10)$$

In this case, from Eq. 8.1 to 8.3, we find $\Delta \geq \omega \geq E_0 - \epsilon_0$. Because of $E_0 - \epsilon_0 \leq 0$, $\omega$ can be either positive or negative dependent on the relative importance of $\rho_l(x)$ and $\rho_c(x)$ (see below for details). And $\sigma$ is uniquely determined by the distribution width of $\rho_l(x)$. Accordingly, the fragility becomes

$$m = \left.\frac{d \log(\tau_\alpha)}{d(T_g/T)}\right|_{T=T_g} = \frac{\omega}{k_B T_g} + 2\left(\frac{\sigma}{k_B T_g}\right)^2. \qquad (11)$$

Eq. 10 and 11 will be used in the following discussions. Hereafter, we call $\omega$ and $\sigma$ the effective barrier and the effective width of the Gaussian distribution, respectively.

According to Eq. 11, the fragility can be well interpreted within the PEL perspective. If $\rho_l(x)$ is a sharp and narrow distribution, i.e., the PEL is flat, the contribution from $\rho_c(x)$ may be neglected, and a strong glass is expected. In this case, the effective barrier $\omega$ will be positive. Conversely, if $\rho_l(x)$ is a very wide and flat distribution, i.e., a rough PEL, $\rho_c(x)$ begins to play a more and more important role, and the effective barrier $\omega$ tends to be negative, then a fragile glass is expected. The above correspondence between PEL roughness and fragility has been confirmed in previous simulations.[29, 41, 45, 63] The shapes of $\rho_l(x)$ and $\rho_c(x)$ for the strong and fragile glass are sketched in the lower panel of Fig. 1.

A quadratic dependence of $\tau_\alpha$ on temperature, similar to the form of Eq.10, was discussed in previous papers, e.g., the Gaussian model and ECG model.[9, 13] However, in the Gaussian model $\omega$ refers to the average barrier, which must be positive. By a simple mathematical transformation, the ECG equation $ln\left(\frac{\tau}{\tau_0^*}\right) = J^2 \left(\frac{1}{T} - \frac{1}{T_0}\right)^2$ can also be deduced from Eq. 10. However, to obtain the ECG equation, a negative $\omega$ must be assumed. If the Eq. 10 is used to fit the AP, a positive $\omega$ will be obtained for typical strong glasses, while $\omega$ becomes a negative number for typical fragile glasses (see below). Obviously, the ECG equation can only be adopted for fragile glasses. In contrast, the Gaussian model describes strong glasses well, but not for typical fragile glasses. In the present model, $\omega$ can be either negative or positive, and has its own physical meaning.

(2) **Comparing to molecular dynamics and experimental data**

Tuning the PEL will be the most direct test of the current model. Recently we have proposed a MD method to adjust the PEL.[66] The advantage of this method is that it only changes the height or distribution of barriers in a PEL, but the number and position of minima in phase space are kept unchanged. Here, this method has been employed to modify the PEL of two nano glasses ($Al_{43}$ and $Al_{46}$ clusters). The relevant computational details are presented in Supplemental Materials. In this method there is a key parameter ($\varepsilon \leq 0$) with $\varepsilon = 0$ meaning no adjustment, and the lower $\varepsilon$ is, the higher the barrier. It is easy to understand that a higher barrier will result in a broader distribution of $\rho_l(x)$, but have less effect on $\rho_c(x)$.

The self-intermediate scattering function (SISF) provides rich information on structural relaxations, which is shown in Fig. 2 for two selected temperatures. $\tau_\alpha$ is given by the intersection of SISFs and the dashed line ($e^{-1}$). It can be seen that, as the PEL becomes rougher ($\varepsilon$ being lower), the structural relaxation does show a significant slowdown, and such a slowdown is more pronounced at low temperatures. This is easy to understand, because the lower $\varepsilon$ is, the higher barrier is.

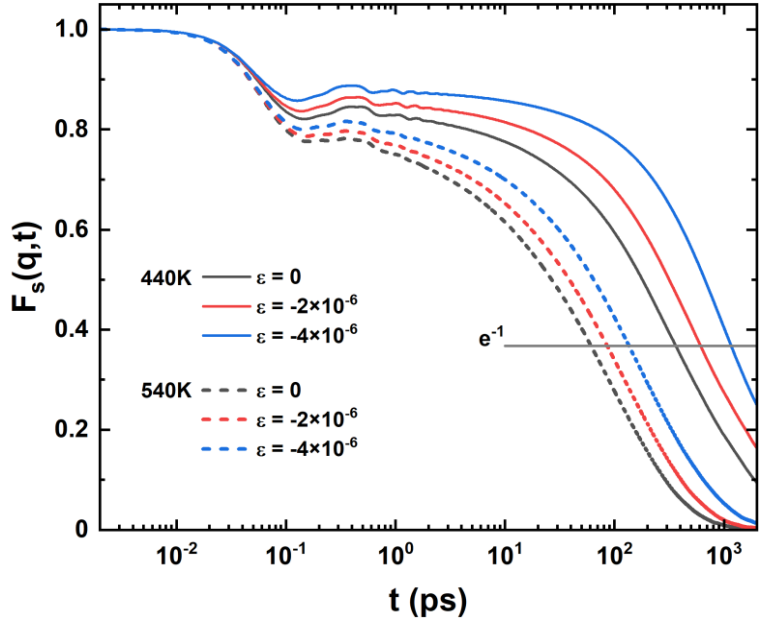

**Figure 2. Self-intermediate scattering function of $Al_{46}$ at 440K and 540K for different $\varepsilon$. As $\varepsilon$ lowers, the structural relaxation time increases significantly.**

The AP of $Al_{43}$ and $Al_{46}$, in the upper panel of Fig. 3, in which $T_g$ is determined according to $\tau_\alpha$, reaches the order of 100 ps. From this figure, one can see that, $Al_{43}$ can be considered a strong glass former at $\varepsilon = 0$, indicated by the Arrhenius relation between $\tau_\alpha$ and temperature. $Al_{46}$ slightly deviates from the Arrhenius relation even at $\varepsilon = 0$, indicating a less strong behavior. With the decrease of $\varepsilon$, the $\tau_\alpha$-temperature curves begin to deviate from the Arrhenius relation, indicating the system tends to be more fragile. This result is actually the most direct support of both our model and previous speculation, namely that the PEL of a fragile glass is "rougher". Because in our current work, as the PEL is adjusted, only barriers are changed. The $\tau_\alpha$-temperature curves from our MD simulations can be well fitted by Eq. 10, as shown in the upper panel of Fig. 3. Besides our MD data, Eq.10 well describes the $\tau_\alpha$-temperature curves of various glasses perfectly. Here the experimental results are taken from Ref.[67-76]. More information for these experimental works can be found in Supplemental Materials. The lower panel of Fig. 3 presents the selected fitted results of the relaxation time by Eq.10, which is almost perfectly fitted (more fitting results are presented in Supplemental

Materials).

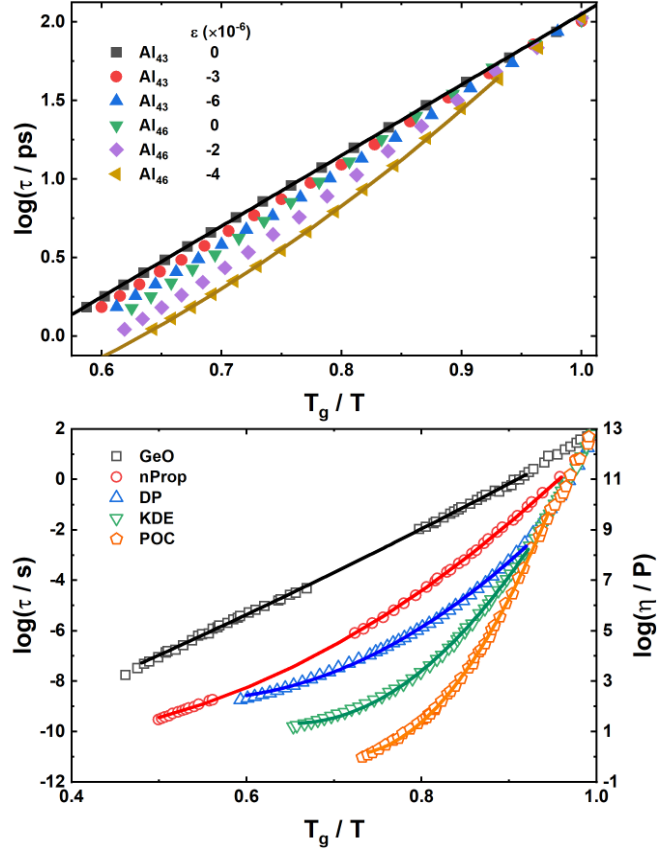

Figure 3. Upper Panel: the Angell plot of $Al_{43}$ and $Al_{46}$ nano glasses. With the decrease of $\varepsilon$, the $\tau_\alpha$-temperature curves gradually deviate from the Arrhenius form, indicating the trend to be fragile. Lower Panel: The Angell plot for various glasses adopted from experimental work. Lines are the fitting results based on Eq. 10. Both the experimental data and the current molecular dynamics data can be well fitted by Eq. 10.

Table 1. The parameters $(\bar{\tau}_0, \omega, \sigma)$ obtained for nano glasses by fitting MD data with Eq. 10.

| System | $\varepsilon (\times 10^{-6})$ | $\ln \bar{\tau}_0$ | $\omega$ (eV) | $\sigma^2 (\times 10^{-2} eV^2)$ |
|---|---|---|---|---|
| | 0 | -5.65 | 0.42 | 0 |
| $Al_{43}$ | -3 | -5.08 | 0.34 | 0.25 |
| | -6 | -4.60 | 0.27 | 0.51 |
| | 0 | -5.91 | 0.38 | 0.36 |
| $Al_{46}$ | -2 | -4.64 | 0.20 | 1.02 |
| | -4 | -2.71 | -0.15 | 2.2 |

The fitting parameters ($\bar{\tau}_0, \omega, \sigma$) for Al$_{43}$ and Al$_{46}$ are listed in Table 1, from which it is seen that our model gives a clear physical description of APs. With decreasing $\varepsilon$, $\omega$ decreases and $\sigma$ increases accordingly. Such a trend is in line with our expectations, since the PEL becomes rougher as $\varepsilon$ decreases. Correspondingly, the system becomes more and more fragile, the effective barrier $\omega$ does not increase but decreases. This indicates that $\rho_c(x)$ begins to play a non-negligible role. At the same time, $\sigma$ grows larger. As discussed above, $\sigma$ is a parameter positively correlated with the effective width of $\rho_c(x)$. The manipulation of the PEL enlarges the difference between barriers, and therefore leads to the increase of $\sigma$. Similar trends between $\omega$ and $\sigma$ are also observed in other glasses (See Table S1 in Supplemental Materials). The above discussion gives us a strong hint that there could be a special relationship between $\omega$ and $\sigma$.

By fitting various glass-forming liquids from polymers, metal alloys and molten salts, and from nano to bulk glasses, we find a simple linear relationship between $\omega$ and $\sigma^2$, as shown in Fig. 4. In this figure, the current MD results for the binary Lennard-Jones system (black open circle) and nano glasses (red open circles) are also included. From this figure, one can see that $\sigma^2$ increases with the decrease of $\omega$. More importantly, $\omega$ and $\sigma^2$ exhibit a linear relationship. All glasses are distributed in a very narrow band, in which fragile and strong glasses locate in the upper left and lower right regions, respectively, in line with our theoretical expectation discussed above.

The narrow-band distribution in Fig. 4 implies that glasses or glass transitions may only occur for a restricted category of PELs, namely within a certain range of $\omega$ and $\sigma^2$. It is not too difficult to argue that far away from this band the glassy state would not exist. Considering two parts that contribute to it, i.e., $\omega_l$ and $\omega_c$, $\omega$ shall has both a maximum and a minimum. On one hand, a large value of $\omega$ corresponds to high local barriers $\omega_l$, which would prevent the system from relaxing towards lower MBs. On the other hand, remember $\omega \geq E_0 - \epsilon_0$ (the difference between the ground-state energy of systems and the zero-temperature energy of the glass), $\omega$ cannot be infinitely small, otherwise the system would be unstable. In fact, although the glassy state is metastable, it could not have a very high energy compared to the ground state energy. What's more, in the $\omega$-$\sigma^2$ plot, we find another intriguing feature, namely, the typical fragile and strong (or weak fragile) glasses are well separated. Thus we can see that the so-called fragile and strong glasses are really different from the perspective of PELs.

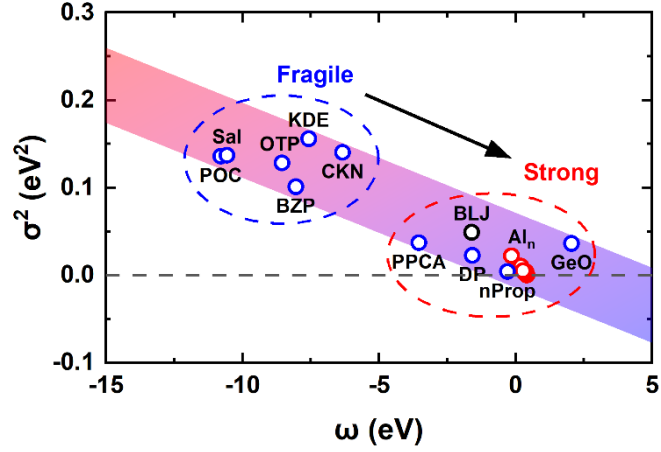

**Figure 4:** $\sigma^2$ **as a function of** $\omega$ **for various glasses.** $\omega$ **and** $\sigma^2$ **seem to have a simple linear relationship. All glasses are distributed in a narrow band. The purple area marks this band, in which the direction of the arrow shows the trend of fragility from fragile towards strong.**

## Conclusion

We have established a model for $\alpha$-relaxation based on the potential energy landscape perspective. We find that the Angell plot can be well explained as long as the barrier distribution is reasonably described. The key issue in our model is to realize that the total distribution of barriers is in two parts, namely a local Gaussian-like distribution, and an extended power-law distribution. Our model shows that a fragile glass tends to have a wide, Gaussian-like distribution, while a strong glass tends to have a narrow Gaussian-like distribution. By comparing various glass systems, we find that these glasses are distributed in a very narrow band determined by two parameters, the square of the effective width of the Gaussian-like distribution and the effective barrier.

**Acknowledgements:** Project supported by the National Natural Science Foundation of China (Grant No. 12188101,11874148). The computations were supported by ECNU Public Platform for Innovation.

# Supplemental Materials to "The Angell Plot from the Potential Energy Landscape Perspective"

D. M. Zhang[a], D. Y. Sun[a,b,c] and X. G. Gong[b,c]

[a]Key Laboratory for Computational Physical Sciences (MOE), Institute of Computational Physics, Fudan University, Shanghai 200433, China
[b]Engineering Research Center for Nanophotonics & Advanced Instrument (MOE), School of Physics and Electronic Science, East China Normal University, 200241 Shanghai, China
[c]Shanghai Qi Zhi Institution, Shanghai 200030, P. R. China

## I. Mathematical deduction for the contribution from $\rho_l(x)$

Considering $f(x) = B_l(x - \Delta)^n \geq 0.0$ and $\rho_l(x) = R_l e^{-f(x)}$, where $B_l$, $R_l$, $\Delta$, and $n$ are constants larger than zero. Defining $F(x) = -\beta x - f(x)$, then $\rho_l(x)e^{-\beta x} = e^{-\beta x - f(x)} = e^{F(x)}$ should be a fast-decayed function of $x$. Namely it is important only in the vicinity of the maximum of $F(x)$. Thus, it is reasonable to replace $F(x)$ by its Taylor expansion,

$$F(x) \cong -\beta \bar{x} - f(\bar{x}) - \frac{1}{2} f''(\bar{x})(x - \bar{x})^2 + O(3).$$

where $\bar{x}$ is determined by $\left.\frac{\partial F}{\partial x}\right|_{\bar{x}} = 0$. Then,

$$\frac{1}{\tau_l} = \frac{R_l}{\tau_0} \int e^{-\beta \bar{x} - f(\bar{x}) - \frac{1}{2} f''(\bar{x})(x - \bar{x})^2 + O(3)} \approx \frac{R_l}{\tau_0} \sqrt{\frac{2\pi}{f''(\bar{x})}} e^{-\beta \bar{x} - f(\bar{x})} \qquad \text{(SI-1)}$$

Now taking $f(x) = B_l(x - \Delta)^n$, we have,

$$\bar{x} = \Delta - \left(\frac{\beta}{nB_l}\right)^{\frac{1}{n-1}}$$

$$f''(\bar{x}) = n(n-1)B_l \left(\frac{\beta}{nB_l}\right)^{\frac{n-2}{n-1}}$$

$$f(\bar{x}) = B_l \left(\frac{\beta}{nB_l}\right)^{\frac{n}{n-1}}$$

After simple mathematical simplification, Eq. SI-1 becomes,

$$\frac{1}{\tau_l} \approx \frac{R_l}{\tau_0} \sqrt{\frac{2\pi}{f''(\bar{x})}} e^{-\frac{\Delta}{k_B T} + (n-1)B_l \left(\frac{1}{nB_l k_B T}\right)^{\frac{n}{n-1}}} \qquad \text{(SI-2)}$$

In order to make $f(x)$ physically meaningful, two constraints are needed. First, the second derivative of $f(x)$ ($f''(\bar{x})$) must be positive, it requires $n > 1$. This, of course, is consistent with that $\rho(x)$ must decay faster than $e^{\beta x}$ for larger $x$. Second, $f''(\bar{x})$ must increase with $T$, otherwise the fluctuation of $\tau_l$ will reach the largest at zero temperature, which obviously makes no sense. This condition requires $n \geq 2$.

## II. Mathematical deduction for the contribution from $\rho_c(x)$

Considering $\rho_c(x) = R_c \left(\frac{x+E_i-E_0}{\mathbb{N}}\right)^{B_c N} = R_c e^{g(x)}$ with $g(x) = B_c N ln\left(\frac{x+E_i-E_0}{\mathbb{N}}\right)$, where $\mathbb{N}, R_c, B_c$ and $E_0$ are constants for a given glass, and $E_i \geq E_0$. Defining $G(x) = -\beta x + g(x)$, then $\rho_c(x)e^{-\beta x} = e^{-\beta x + g(x)} = e^{G(x)}$ should be a fast-decayed function of $x$. Namely it is important only in the vicinity of the maximum of $G(x)$. Thus, it is reasonable to replace $G(x)$ by its Taylor expansion,

$$G(x) \cong -\beta \bar{x} + g(\bar{x}) + \frac{1}{2} g''(\bar{x})(x-\bar{x})^2 + O(3),$$

where $\bar{x}$ is determined by $\left.\frac{\partial G}{\partial x}\right|_{\bar{x}} = 0$, thus

$$\bar{x} = E_0 - E_i + \frac{B_c N}{\beta}$$

and

$$g(\bar{x}) = B_c N ln\left(\frac{B_c N}{\mathbb{N}\beta}\right)$$

$$g''(\bar{x}) = -NB_c \left(\frac{B_c N}{\beta}\right)^{-2}$$

$$\frac{1}{\tau_c} = \frac{R_c}{\tau_0} \int e^{-\beta \bar{x} + g(\bar{x}) + \frac{1}{2}g''(\bar{x})(x-\bar{x})^2 + O(3)} \approx \frac{R_c}{\tau_0} \sqrt{\frac{2\pi}{|g''(\bar{x})|}} e^{-\beta \bar{x} + g(\bar{x})} \quad \text{(SII- 1)}$$

$$-\beta \bar{x} + g(\bar{x}) = \beta(E_i - E_0) + B_c N + B_c N ln\left(\frac{B_c N}{\mathbb{N}\beta}\right)$$

Then

$$\frac{1}{\tau_c} \approx \frac{R_c}{\tau_0} \sqrt{\frac{2\pi}{|g''(\bar{x})|}} e^{\frac{E_i}{k_B T} - \frac{E_0}{k_B T} + B_c N + B_c N ln\left(\frac{B_c N}{\mathbb{N}\beta}\right)} \quad \text{(SII- 2)}$$

To make $g(x)$ have physical meaning, similar to that for $g(x)$, two constraints are needed. The second derivative of $g(x)$ ($g''(\bar{x})$) must be negative, and decrease with $T$, which are automatically satisfied.

## III. Computational Methods

The thermodynamics behavior of two aluminum nanoclusters ($Al_{43}$ and $Al_{46}$) were studied, and the empirical potential was adopted to describe the atomic interaction of $Al^{18}$. The constant temperature MD method without any boundary conditions was used in the calculations. At each temperature of interest, a 10ns simulation for initial relaxation is performed, followed by a 6μs simulation. Previous studies had shown that both $Al_{43}$ and $Al_{46}$ have a disordered structure at its ground state, and melt or solidifies with a typical glass-like transition, namely the continuous change in energy and volume[19-21].

In order to manipulate PELs, the method developed by some of us was used. Here the key points were listed as following: the high-dimensional PELs are the function of the total potential energy of systems (**Note: the total potential energy is not the interatomic potential**). Suppose $\varphi$ being the total potential energy function, a new total potential energy $\varphi^*$ is defined as,

$$\varphi^* = \begin{cases} \varphi + \varepsilon(\varphi - \varphi_0)^m & \text{for } \varphi \leq \varphi_0 \\ \varphi & \text{for } \varphi > \varphi_0 \end{cases}, \quad \text{(SIII-1)}$$

where $\varepsilon$, $\varphi_0$, and $m$ are adjustable parameters. It can be easily demonstrated that, by proper choice of $\varepsilon$, $\varphi_0$ and $m$, the new PEL determined by $\varphi^*$ has the adjustable barriers but the same topologic structure as the one determined by $\varphi$. Namely by adjusting PELs, both the number and position of extreme points in phase space is unchanged, but only the height of barriers or the roughness of PELs is modified, which is a remarkable feature of our method. In current studies, for $Al_{43}$ ($Al_{46}$), $\varphi_0$ and $m$ were $-2.58 \times 43(46)$ eV and 6, respectively. And $\varepsilon$ was chosen in the range of $[-6 \times 10^{-6}, 0]$. Since here $\varepsilon$ is negative, the PEL becomes steeper as $\varepsilon$ is decreased, in other word, the fluctuation of barriers increases.

To obtain the structural relaxation time (the so-called $\alpha$-relaxation time, $\tau_\alpha$), the self-intermediate scattering function (SISF), which reveals a stretched time evolution of structure, is calculated by

$$F_s(q,t) = N^{-1} \langle \sum_{i=1}^{N} \exp\{i\boldsymbol{q} \cdot [\boldsymbol{r}_i(0) - \boldsymbol{r}_i(t)]\} \rangle, \quad \text{(SIII-2)}$$

where $N$ is the number of atoms in nanoclusters, $\langle \rangle$ denotes the ensemble average,

and $q$ is the wave vector. In this work, $|q| = q_{max} = 2.84 Å^{-1}$, i.e., the position of the main peak in the static structure factor $S(q)$[23, 24]. From SISF, one can obtain the structural relaxation time $\tau_\alpha$, which is usually defined as $F_s(q, t = \tau_\alpha) = e^{-1}$.

## IV. Fitting to MD and Experimental Data

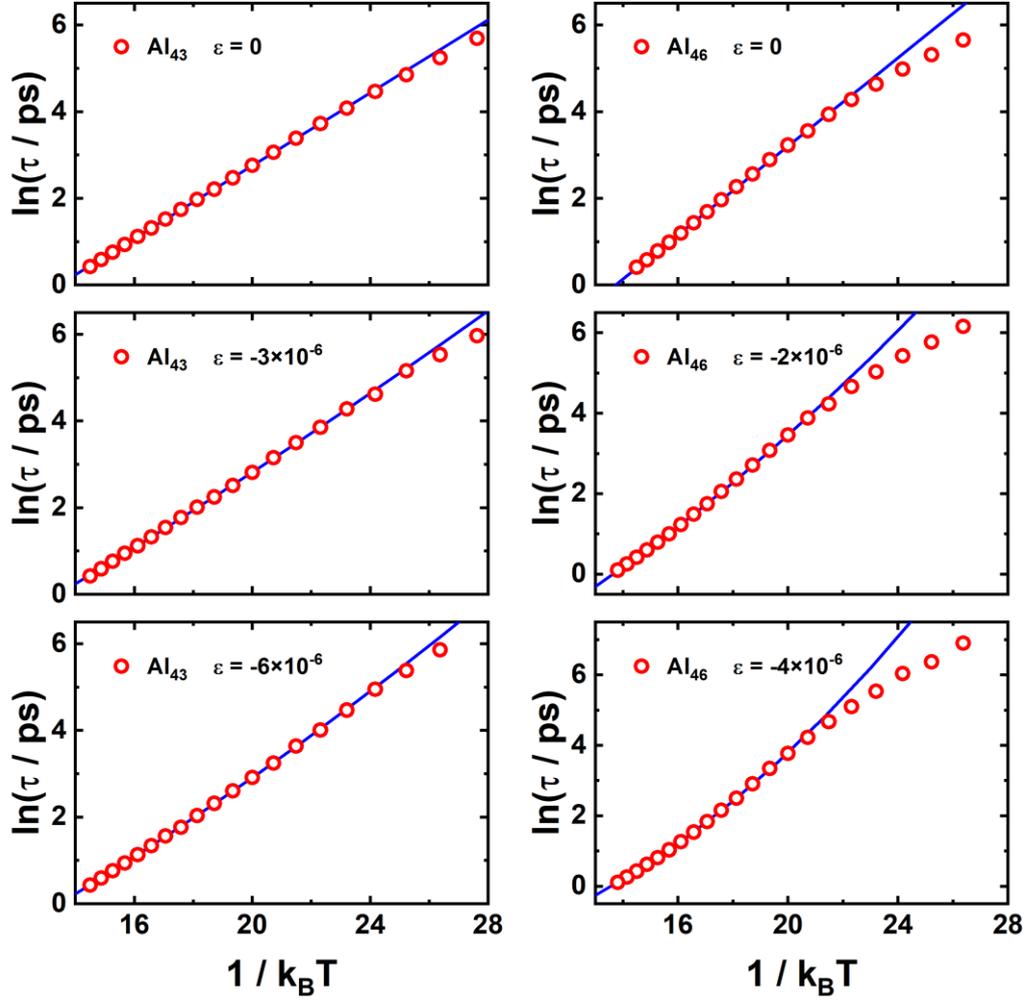

Figure S1. Fitting MD results with Eq. 10.

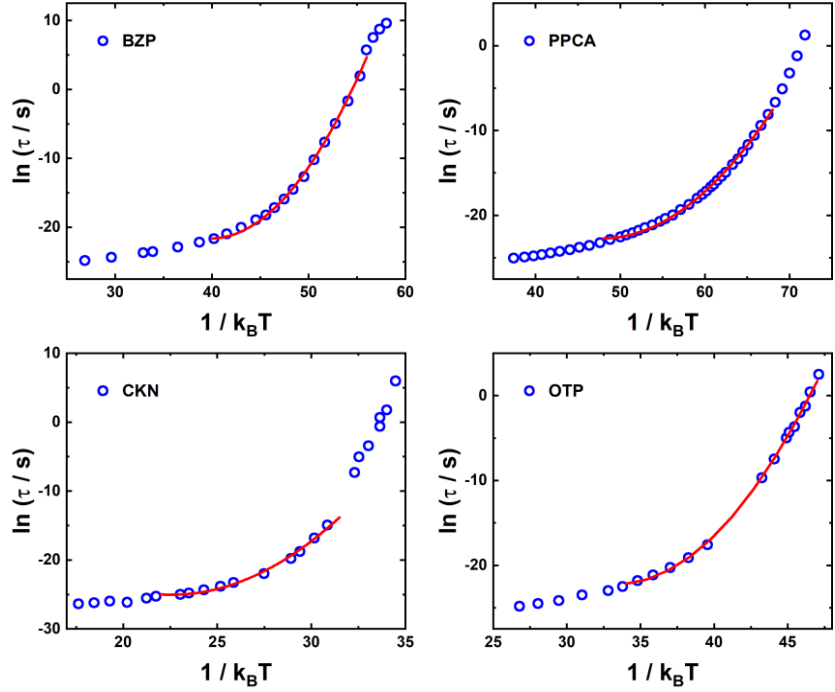

Figure S2. Fitting experimental results with Eq. 10.

Table 1. The parameters ($\bar{\tau}_0, \omega, \sigma$) obtained by fitting experimental results with Eq.10

| System | Full name | ln($\tau_0$) | ln($\eta_0$) | $\omega$ | $\sigma^2(\times 10^{-2})$ |
|---|---|---|---|---|---|
| GeO | GeO$_2$ | | -7.01 | 2.05 | 3.61 |
| nProp | n-propanol | -18.41 | | -0.29 | 0.40 |
| DP | dibutyl-phthalate | 7.70 | | -1.58 | 2.26 |
| KDE | cresolphthalein-dimethylether | 69.69 | | -7.57 | 15.58 |
| Sal | salol | 182.48 | | -10.56 | 13.68 |
| OTP | o-terphenyl | 120.32 | | -8.55 | 12.83 |
| PPCA | propylene carbonate | 61.79 | | -3.54 | 3.70 |
| CKN | Ca-K-NO$_3$ | 46.34 | | -6.33 | 14.03 |
| BZP | benzophenone | 138.75 | | -8.05 | 10.10 |
| POC | α-phenyl-o-cresol | | 214.37 | -10.79 | 13.58 |

In many glass formers there exists such a fragile-strong crossover, i.e., at low temperature region, the temperature dependence of structural relaxation time and viscosity shows a non-Arrhenius to Arrhenius crossover, which is thought to be connected to the breakdown of ergodicity. Due to the broken ergodicity, the system no longer "sees" the complete distribution of barriers, so the relationship does not hold below the crossover temperature, see, for example, fitting results of $Al_{46}$, BZP and CKN.

## References for Experimental Data: